\RequirePackage{lineno}

\documentclass[twoside,slac_one]{revtex4}
\usepackage{graphicx}
\usepackage{fancyhdr}
\usepackage{amsmath} 
\usepackage{bm}
\usepackage{amsxtra}
\usepackage{amssymb}
\usepackage{amsthm}
\usepackage{latexsym}
\usepackage{lscape}

\usepackage{enumerate}

\def\Lb{$\Lambda_b$}			 
\def\L0{$\Lambda$}			 
\def\aL0{$\overline{\Lambda}$}      

\def\Lbch{$\Lambda_b \rightarrow J/\psi \Lambda$}  
\def\Bch{$B^0 \rightarrow J/\psi K^0_S$}  

\def\B0{$B^0$}			 
\def\aB0{$\overline{B^0}$}      
\def\K0{$K^0_S$}			 

\def\bbar{$\overline{b}$}               
\def\jpsi{$J/\psi$}                     
\def\ifb{fb$^{-1}$}                     
\def\deg{$^\circ$}                      

\def\D0{D0}  

\begin{document}

\title{Measurement of the production fraction times branching fraction $\boldsymbol{ f(b\to\Lambda_{b})\times \mathcal{B}(\Lambda_{b}\to J/\psi \Lambda)}$}

%

\author{I. Heredia-De La Cruz}
\affiliation{Department of Physics, CINVESTAV, Mexico City, Mexico \hfill \; \\
\textit{On behalf of the \D0 Collaboration} }

\begin{abstract}
A new measurement of the $b \to \Lambda_b$ production fraction multiplied by the \Lbch~branching fraction was performed by the \D0 experiment using 6.1 fb$^{-1}$ of $p\bar{p}$ collisions at $\sqrt{s}=1.96$ TeV. The result of this measurement, $f(b\to\Lambda_{b})\cdot\mathcal{B}(\Lambda_{b}\to J/\psi \Lambda) = \left [6.01\pm 0.60\mbox{ (stat.)}  \pm 0.58\mbox{ (syst.)} \pm  0.28 \mbox{ (PDG)} \right ] \times 10^{-5}$, represents an improvement in precision by about a factor of three with respect to the current  world average. We give an estimate of $\mathcal{B}(\Lambda_{b}\to J/\psi \Lambda)$, which takes into account  correlations among the different $b$-hadron production fractions and other weakly decaying baryons.
\end{abstract}

\maketitle

\thispagestyle{fancy}


\section{Introduction}

Until recently, the only particle collider capable of producing $b$ baryons was the Fermilab Tevatron Collider. Due to their relatively heavy mass, their production is suppressed with respect to the more favored $B$ mesons, and even for the lightest and most copiously produced $b$ baryon, the $\Lambda_{b}(udb)$, only a few decay channels and properties have been studied. In particular, the uncertainties on \Lb~branching fractions are on the order of $\sim$($30$--$60$)\%. With the full datasets of the \D0 and CDF experiments at Fermilab and the excellent performance of the Large Hadron Collider (LHC) and the experiments at CERN, it will be also possible to study more precisely important effects on $b$ baryons, such as polarization, CP and T violation.

The \D0 Collaboration reports a measurement of the production fraction multiplied by the branching fraction of the \Lbch~decay relative to that of the decay \Bch~\cite{D0BrLb}, 
\begin{linenomath*}
\begin{equation}\label{eq:sigmareldef}
\sigma_{\text{rel}} \equiv \frac{f(b\to\Lambda_{b})\cdot\mathcal{B}(\Lambda_{b}\to J/\psi \Lambda)}{f(b\to B^{0}) \cdot\mathcal{B}(B^{0}\to J/\psi K^{0}_{S})}.
\end{equation}
\end{linenomath*}
The estimation of $f(b\to\Lambda_{b})\cdot\mathcal{B}(\Lambda_{b}\to J/\psi \Lambda)$ is provided based on the best value of $f(b\to B^{0}) \cdot\mathcal{B}(B^{0}\to J/\psi K^{0}_{S})$~\cite{PDG}. A description of this analysis is given in the following sections. Finally, we give our estimate of $\mathcal{B}(\Lambda_{b}\to J/\psi \Lambda)$.

\section{Experimental and theoretical status of \boldsymbol{$\mathcal{B}(\Lambda_{b}\to J/\psi \Lambda)$}} \label{sec:ExpTheo}
The last measurement of $\sigma_{\text{rel}}$ was performed by the CDF experiment~\cite{CDFBrLbrecent} with only $7.8 \pm 3.4$ \Lb~signal candidates. They found 
\begin{linenomath*}
\begin{equation}
\label{eq:sigmarelWA}
\sigma_{\text{rel}}^{\text{W.A.}} = 0.27 \pm 0.12 \text{ (stat.)} \pm 0.05 \text{ (syst.)},
\end{equation}
\end{linenomath*}
where W.A. stands for World Average. Based on this result the Particle Data Group (PDG)~\cite{PDG} reports
\begin{linenomath*}
\begin{eqnarray}
\label{eq:fBLWA}
f(b\to\Lambda_{b})\cdot\mathcal{B}(\Lambda_{b}\to J/\psi \Lambda) = (4.7\pm 2.3)\times 10^{-5}.
\end{eqnarray}
\end{linenomath*}
With more statistics and improved simulation of the processes involved and the experimental environment, this measurement can be greatly improved. It is important to mention that the dominant systematic uncertainty on this measurement is the unknown \Lb~polarization (section~\ref{subsubsec:Lbpol}). 

On the other hand, there are several theoretical predictions of this branching fraction. For example, Ref.~\cite{PQCD} uses perturbative QCD to find $\mathcal{B}(\Lambda_{b}\to J/\psi \Lambda) \sim (1.65-5.27) \times 10^{-4}$. The same branching fraction is calculated in the framework of the factorization hypotheses~\cite{Fayyazuddin}, using relativistic~\cite{Rel1,Rel2,Rel3} and non-relativistic~\cite{NoRel1,NoRel2} quark models, and ranges from $\sim (1.1-6.1) \times 10^{-4}$.

\section{Detector} \label{sec:Detector}
The \D0 detector is described in detail in Ref.~\cite{d0det}. In general, in order to study $B$ decays, the most relevant components are the central tracking system and the muon spectrometer. The \D0 central tracking system is composed of a silicon microstrip tracker (SMT) and a central fiber tracker (CFT) covering the pseudorapidity region $|\eta|<3.0$ and $|\eta|<2.0$, respectively, where $\eta \equiv -\ln[\tan(\theta/2)]$ and $\theta$ is the polar angle. They provide the ability to reconstruct charged tracks and vertices in a highly busy environment (typically more than 100 charged tracks), and a surrounding 2~T superconducting solenoid allows precise measurements of the transverse momentum ($p_T$) of the particles. The muon spectrometer consist of three layers of drift tubes and scintillator trigger counters, one located in front and two after 1.8~T iron toroids, and covering up to $|\eta|<2.2$.

\section{Data sample and event reconstruction}
This analysis uses an integrated luminosity of about $6.1$ \ifb~recorded by the \D0 detector from 2002--2009 at $\sqrt{s}=1.96$ TeV. The data sample consists of events that satisfy single muon or dimuon triggers. 

\subsection{Event reconstruction}
The decay topology of \Lbch~and \Bch~is shown in Fig.~\ref{fig:topology}.\footnote{ Unless explicitly stated otherwise, the appearance of a specific charge state also implies its charge conjugate.} The strategy to search for these decays is the following: 
\begin{enumerate}[(i)]
 \item Look for events with two oppositely charged reconstructed muons, forming a common vertex, and with invariant mass $M(\mu^+\mu^-)$ in the range $2.8-3.35$~GeV/$c^2$. Muons are identified by matching tracks reconstructed in the central tracking system with  track segments in the muon spectrometer.
 \item Search for pairs of oppositely charged tracks with a common vertex in those events satisfying the dimuon selection. For the \L0~reconstruction, Monte Carlo (MC) studies support that the track with the highest $p_T$ is the proton.
 Events within 1.102 $ < M(p\pi^-) < $ 1.130~GeV/$c^2$  and  0.466 $ < M(\pi^{+}\pi^{-}) < $  0.530 GeV/$c^2$ are selected. 
 \item \Lb~and \B0~candidates are reconstructed by performing a constrained fit to a common vertex for  the \L0~or \K0~candidate (a neutral track which is propagated from the $p\pi^-$ or $\pi^{+}\pi^{-}$ common vertex according to the momentum direction of the $p\pi^-$ or $\pi^{+}\pi^{-}$) and the two muon tracks. In this fit the dimuon mass is constrained to the W.A. $J/\psi$ mass~\cite{PDG}. Events within $5.0 < M(J/\psi \Lambda) < 6.2$~GeV/$c^2$ and $4.8 < M(J/\psi K^0_S) < 5.8$~GeV/$c^2$ are selected.
 \item Finally, the reconstruction algorithm must be able to identify at least one $p\bar{p}$ interaction vertex\footnote{The $p\bar{p}$ interaction vertex is determined by  minimizing a  $\chi^2$ function that depends on all reconstructed tracks in the event and a term that represents the average beam position constraint.}. In case of multiple interaction $p\bar{p}$ vertices in the event, the one closest to the $B$ candidate vertex is tagged as the primary vertex (PV) for this candidate. 
\end{enumerate}

 \begin{figure}
 \includegraphics[scale=0.35]{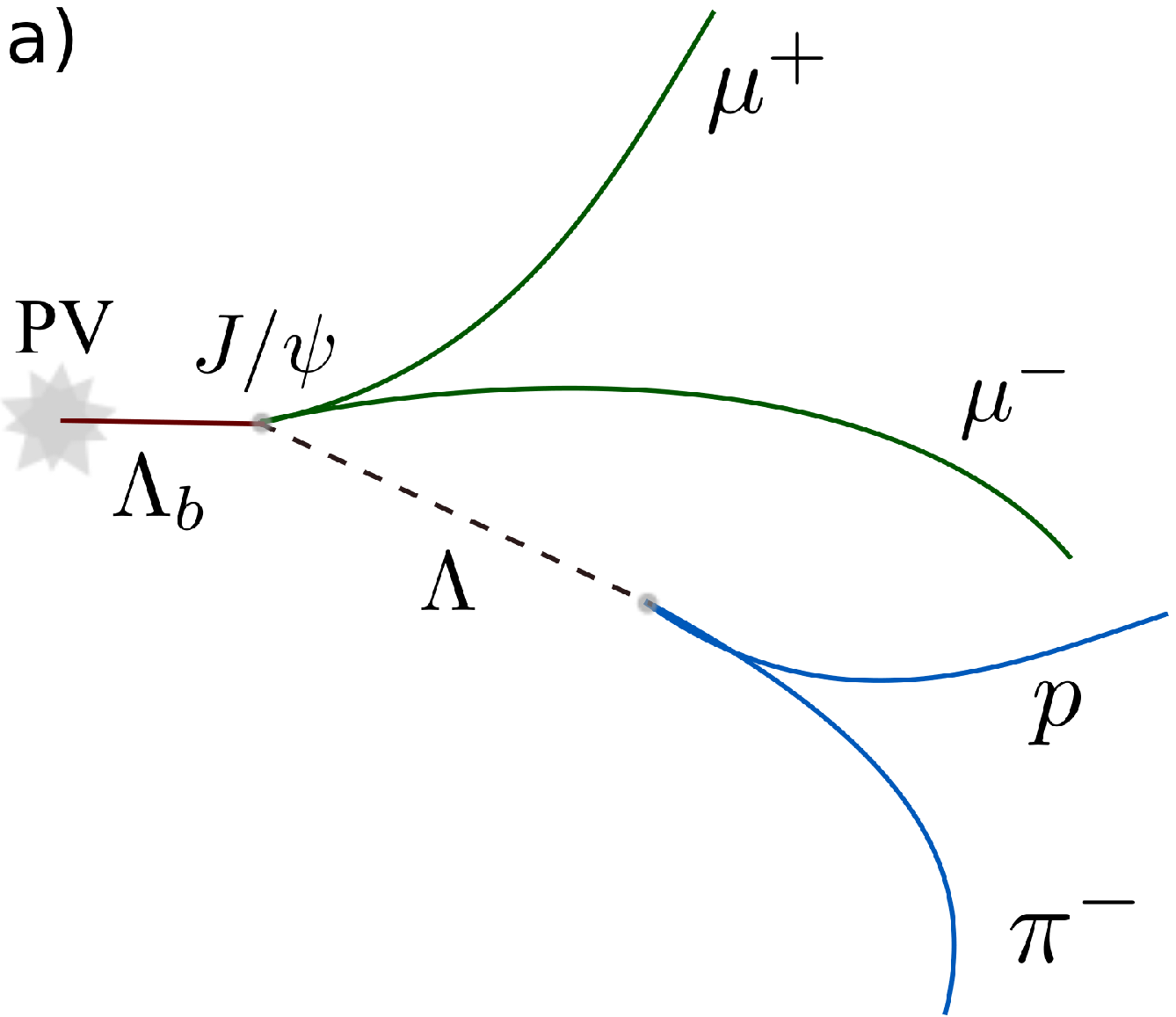} \; \; \; \; \;
 \includegraphics[scale=0.35]{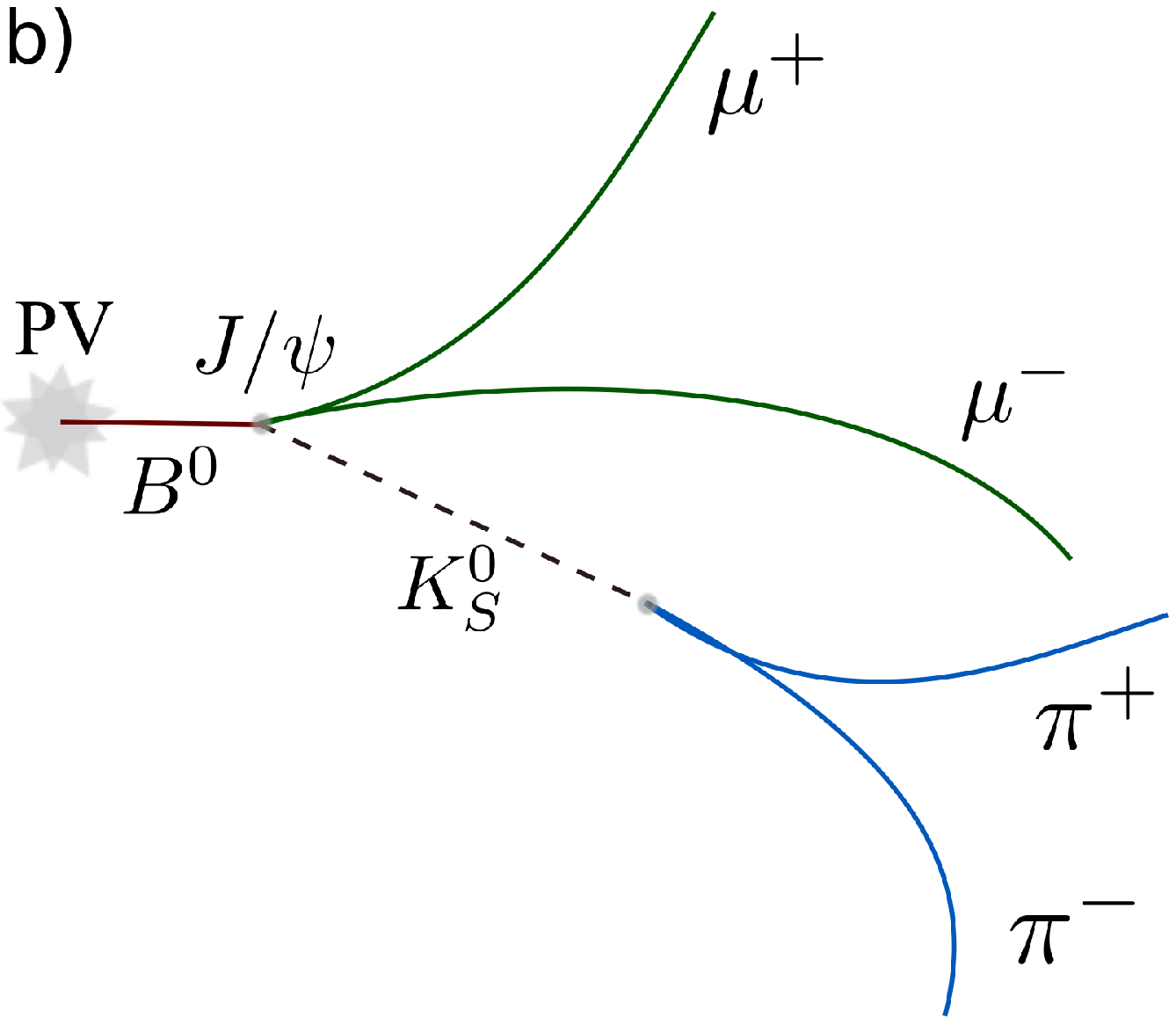}
 \caption{\label{fig:topology} Topology of the decays a) \Lbch~and b) \Bch, with \jpsi~$\to \mu^+\mu^-$, $\Lambda \to p\pi^-$ and $K^0_S \to \pi^+\pi^-$.  }
 \end{figure}

\subsection{Event quality and background suppression} \label{subsec:bkgsup}
Several conditions are imposed on the quality of the reconstructed objects (tracks, vertices and parent particles):

\begin{enumerate}[(i)]
 \setcounter{enumi}{4}
 \item Every muon track must be associated to at least two hits in (both) the SMT and CFT, and satisfy $p_T>2.0$~GeV/$c$ and $|\eta|<2.0$. At least one muon must have segments in the muon system  inside and outside the toroid. 
 \item Since \L0~and \K0~are long-lived particles, they are likely to decay outside the beam pipe (and many of them outside the central tracking system). No detection condition is required in the SMT for the daughter tracks (proton or pion candidates); however, each of them must be detected with at least one hit in the CFT and, in total, they must not have more than two hits in the tracking detectors between the primary vertex and the common two-track vertex. Also, the impact parameter significance (the impact parameter with respect to the primary  vertex divided by its uncertainty) is required to exceed $3$ for both tracks and $4$ for at least one of them. 
 \item All (\Lb, \B0, \L0, \K0~and \jpsi) decay vertices must be well reconstructed, with a $\chi^2$ probability greater than 1\%.
\end{enumerate}

In order to suppress undesirable backgrounds (distributed below or very close to the signal peaks) such as the cross-feed contamination\footnote{The \Lb~sample may be contaminated  with \B0~events that pass the \Lb~selection, or vice versa.\label{crossfeed} } between \Lb~and \B0, and cascade decays of more massive baryons like $\Sigma^{0}\to\Lambda\gamma$ and $\Xi^{0}\to\Lambda\pi^{0}$, it is required that:
\begin{enumerate}[(i)]
 \setcounter{enumi}{7}
  \item Track pairs simultaneously identified as both \L0~and \K0~due to different mass assignments to the same tracks are removed. 
  \item The pointing angle\footnote{To be precise, this is the angle between the $p_T$ of the \L0~and the vector from the \jpsi~vertex to the \L0~decay vertex in the plane perpendicular to the beam direction.} of the \L0~(\K0) track to the \jpsi~vertex in the transverse plane  must not exceed 2.5\deg.
\end{enumerate}

Finally, one can take advantage of the topology and kinematics of these decays in order to determine the final selection criteria. For example, it is easy to get rid of the prompt background (mainly $J/\psi$'s coming from the primary vertex plus random tracks) by applying a minimum cut on the reconstructed decay length of the $B$ particle. Similarly, the long-lived nature of the \L0~and \K0~can be used to suppress combinatorial background. To decide the final selection, MC events are generated for $\Lambda_b \to J/\psi (\mu^+\mu^-)\Lambda$ and $B^0 \to J/\psi (\mu^+\mu^-)K^0_S$ using {\sc pythia}~\cite{Pythia} and  {\sc evtgen}~\cite{EvtGen} for the production and decay simulation, followed by full modeling of the detector response with {\sc geant}~\cite{Geant},  taking into account the effects of multiple interactions at high luminosity by overlaying hits from randomly triggered $p\bar{p}$ collisions on the digitized hits from MC, and event reconstruction as in data. Then the figure of merit $\mathrm{S}=N_{S}/\sqrt{N_{S}+N_{B}}$ is maximized, where $N_{S}$ is the number of signal candidates determined by MC and $N_{B}$ the number of background candidates estimated by using data events in the sidebands of the expected signal. One ends up with the following requirements:
\begin{enumerate}[(i)]
 \setcounter{enumi}{9}
 \item Dimuon candidates must satisfy $p_{T}(\mu^{+}\mu^{-})>3.0$ GeV/$c$. For the  \L0~(\K0), the  $p_T$ must be greater
than 1.6 (1.0) GeV/$c$, the transverse decay length greater than 0.8 (0.4) cm and its significance greater than 4.0 (9.0). For the \Lb~(\B0) candidate, the  $p_T$ must be greater than 5.0 GeV/$c$ and the significance of the proper decay length\footnote{The  proper decay length is defined as $L_{xy}M/p_{T}$, being $p_T$ and $M$ the transverse 
momentum  and mass of the $b$ hadron, respectively, and  $L_{xy}$ the  distance between the primary vertex 
and the $b$ hadron decay vertex in the transverse plane.} is required to be greater than 2.0 (3.0). 
\end{enumerate}
It may be the case that multiple candidates are found in the same event, for which only the one with the best $\chi^2$ probability of the $B$ decay vertex is selected.

The invariant mass distributions of the events satisfying the selection (i -- x) are shown in Fig.~\ref{fig:mass}. An unbinned likelihood fit to each distribution yields  $N_{\Lambda_{b}\to J/\psi\Lambda}=314\pm 29$ and $N_{B^{0}\to J/\psi K^{0}_{S}}=2335\pm 73$ candidates.

 \begin{figure}
 \includegraphics[scale=0.4]{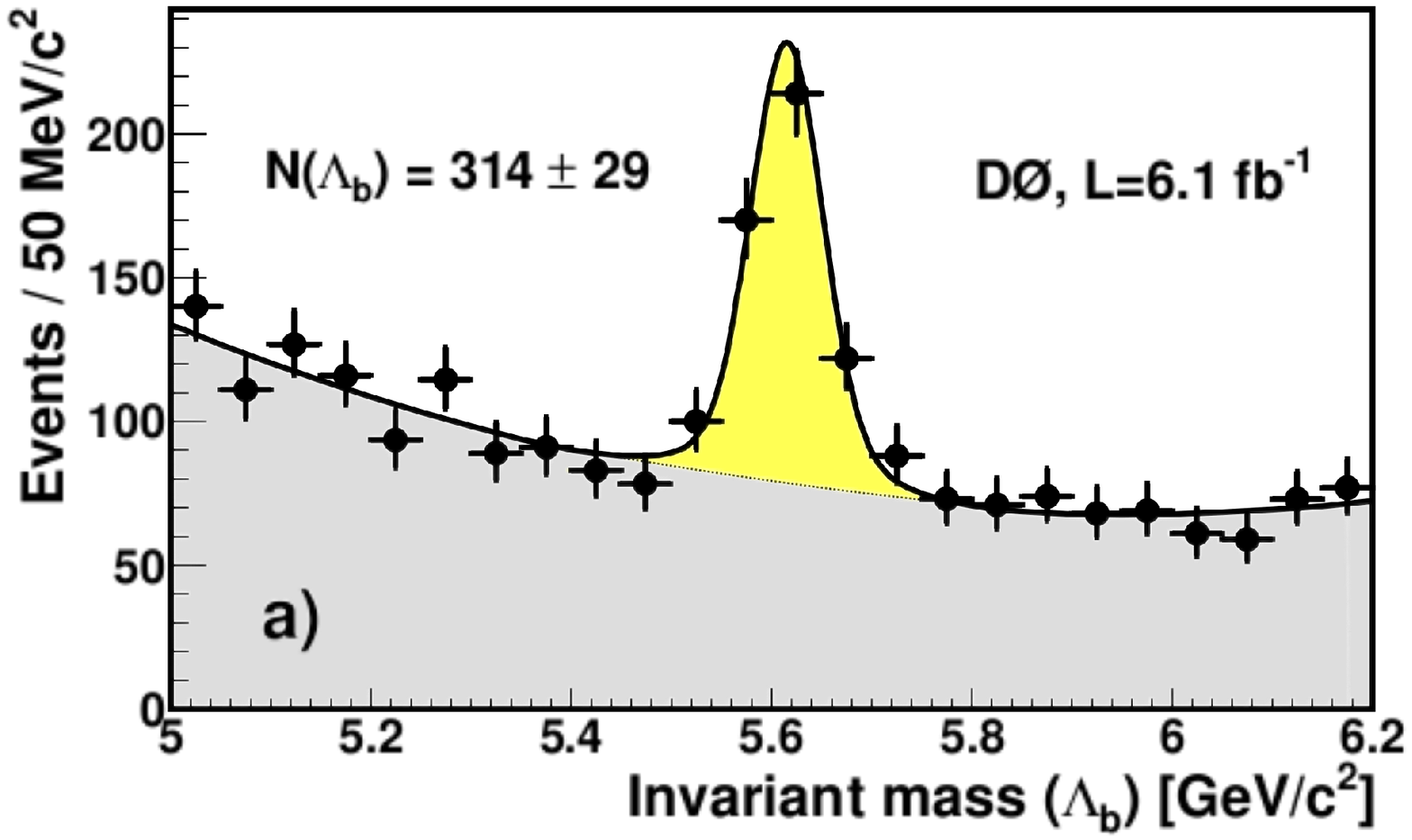}
 \includegraphics[scale=0.4]{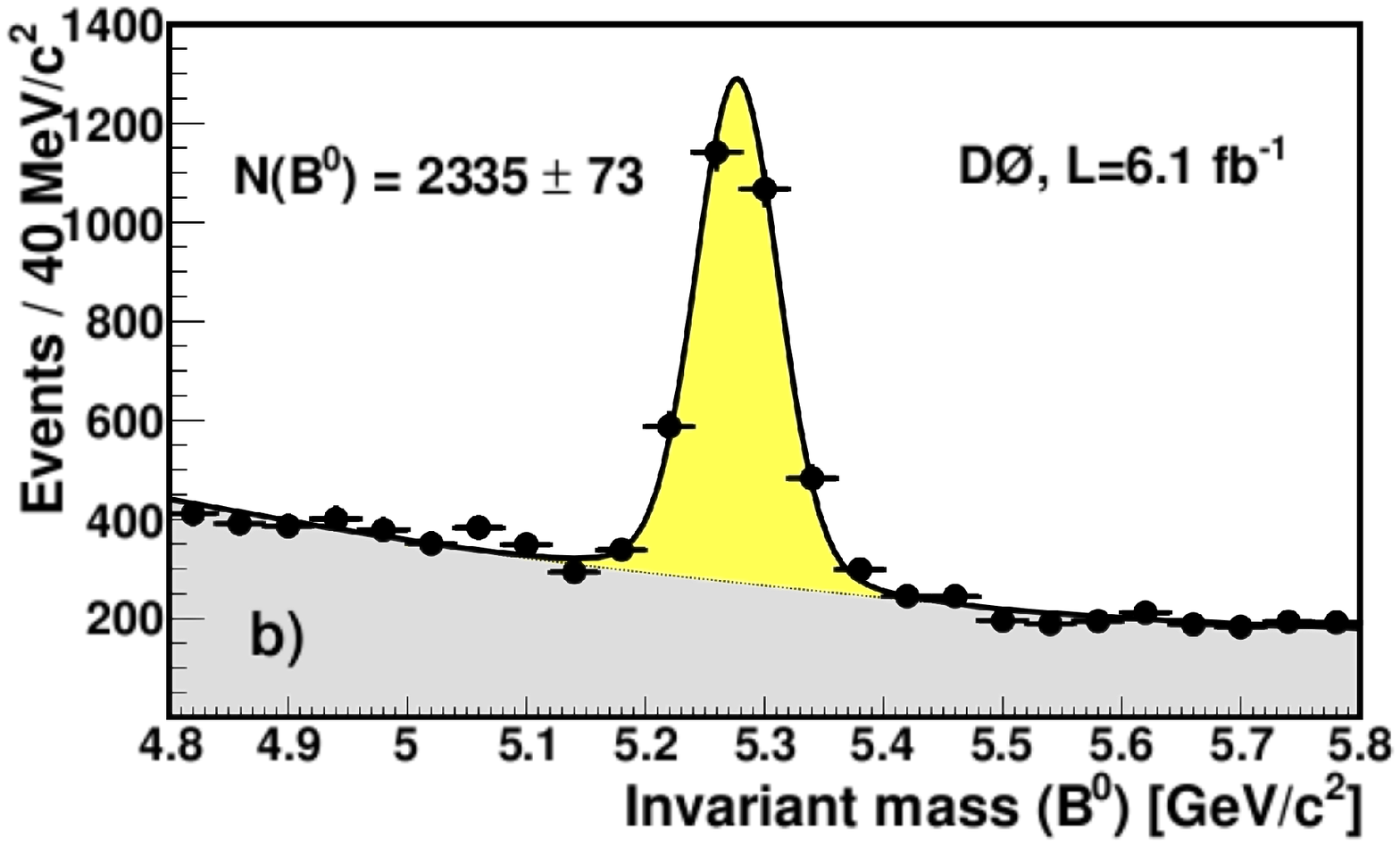}
 \caption{\label{fig:mass} Mass distribution of  a) \Lb~and b)
 \B0~candidates. The background distribution is parametrized as a second order polynomial and the signal distribution as a double Gaussian function.}
 \end{figure}

\section{Branching fraction measurement}
\subsection{The method}
The number of observed \Lb~events (found in the previous section) decaying to $J/\psi\Lambda$, with the \jpsi~going a pair of muons and \L0~to a proton and a pion, is given by
\begin{linenomath*}
\begin{equation}\label{eq:first}
N_{\Lambda_{b}\to J/\psi\Lambda} = N_{prod}\left[\Lambda_b \rightarrow J/\psi(\mu^+ \mu^-)\Lambda(p \pi^-)\right]  \times \epsilon_D\left[\Lambda_b \rightarrow J/\psi(\mu^+ \mu^-)\Lambda(p \pi^-)\right],
\end{equation}
\end{linenomath*}
where the number of decays produced in collisions is
\begin{linenomath*}
\begin{equation}\label{eq:second}
N_{prod}\left[\Lambda_b \rightarrow J/\psi(\mu^+ \mu^-)\Lambda(p \pi^-)\right] =  \mathcal{L} \sigma(p\overline{p} \rightarrow b\overline{b}) f(b\rightarrow\Lambda_b)   
 \mathcal{B}\left(\Lambda_b \rightarrow J/\psi \Lambda\right) \mathcal{B}\left(J/\psi \rightarrow \mu^+ \mu^- \right)  \mathcal{B}\left(\Lambda \rightarrow p \pi^-\right).
\end{equation}
\end{linenomath*}
Here $\mathcal{L}$ is the integrated luminosity and $\sigma(p\overline{p} \rightarrow b\overline{b})$ is the cross-section for the production of  $b$\bbar~quarks. The detection efficiency $\epsilon_D\left[\Lambda_b \rightarrow J/\psi(\mu^+ \mu^-)\Lambda(p \pi^-)\right]$ encompasses acceptance effects as well as detector, trigger and reconstruction efficiencies for this decay. This efficiency is obtained from MC simulation. 

Similar expressions to Eqs.~(\ref{eq:first}) and~(\ref{eq:second}) can be obtained for $B^0 \to J/\psi(\mu^+\mu^-)K^0_S(\pi^+\pi^-)$. Then, it is easy to show that
\begin{linenomath*}
\begin{equation} \label{eq:sigmarel}
\sigma_{\text{rel}}  = \frac{N_{\Lambda_{b}\to J/\psi\Lambda}}{N_{B^{0}\to J/\psi K^{0}_{S}}} \cdot \frac{\mathcal{B}(K^{0}_{S}\to\pi^{+}\pi^{-})}{\mathcal{B}(\Lambda\to p\pi^{-})} \cdot \epsilon_{\text{rel}} ,
\end{equation}
\end{linenomath*}
where $\sigma_{\text{rel}}$ is defined in Eq.~(\ref{eq:sigmareldef}) and the relative detection efficiency $\epsilon_{\text{rel}} \equiv \epsilon_D[B^{0}\to J/\psi K^{0}_{S}]  / \epsilon_D[\Lambda_{b}\to J/\psi\Lambda]$ is determined in the next section. For now, it is important to mention that most systematic and detector effects which are not fully implemented in the simulation (dead channels, trigger effects, pile-up, etc.) will cancel out in this ratio. Quantities such as $b$ quark production, integrated luminosity and (to some extent) selection efficiencies are also canceled in $\sigma_{\text{rel}}$. Hence the importance of choosing a normalization channel topologically equivalent to the decay under study.

\subsection{Detection efficiencies}
In order to determine the detection efficiencies, independent MC samples (different from the samples used to optimize the selection) of \Lb~and \B0~decays are generated, with the same procedure described in section~\ref{subsec:bkgsup}. Important effects such as tracking detector efficiencies  and luminosity dependence with time are incorporated in the simulation for different detector epochs and by the overlay of zero-bias events (triggered solely on the bunch crossing time). The same process for reconstructing and selecting events as in data is strictly followed. All the variables used in the selection are found to be in good agreement between data and MC. A noteworthy example is shown in Fig.~\ref{fig:PDL}, where the proper decay length distribution of \K0~candidates is compared. Although no significant mismodeling was found in the simulation, any residual effect is expected to be reduced in the ratio of detection efficiencies. The relative detection efficiency of \Bch~and \Lbch~decays is found to be
\begin{linenomath*}
\begin{equation} \label{eq:epsrel}
 \epsilon_{rel} = 2.37 \pm 0.05 \text{ (MC stat.)}.
\end{equation}
\end{linenomath*}

\begin{figure}
\includegraphics[scale=0.38]{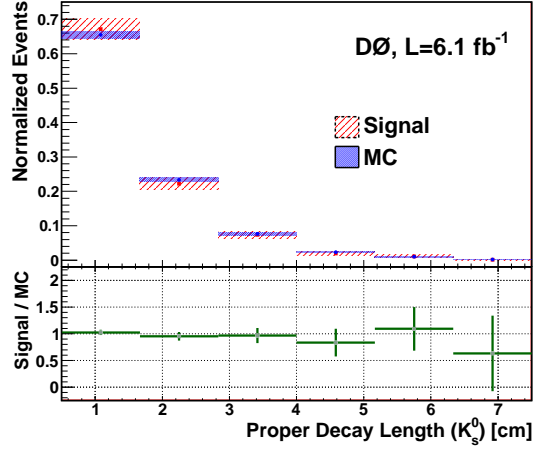}
\caption{\label{fig:PDL} Proper decay length distributions for \K0~candidates reconstructed in the decay \Bch, comparing data and simulation.
}
\end{figure}

Using the number of reconstructed \Lb~and \B0~decays in Fig.~\ref{fig:mass}, $\mathcal{B}(K^{0}_{S}\to\pi^{+}\pi^{-})=0.6920\pm0.0005$
and $\mathcal{B}(\Lambda\to p\pi^{-})=0.639\pm0.005$~\cite{PDG}, we obtain
\begin{linenomath*}
\begin{equation}\label{eq:sigmarelmed}
\sigma_{\text{rel}}=0.345 \pm 0.034 \mbox{ (stat.)} \pm 0.003 \mbox{ (PDG)},
\end{equation}
\end{linenomath*}
where the uncertainty due to inputs from the PDG~\cite{PDG} have been separated.

\subsection{Systematic uncertainties}
The sources of systematic uncertainties in the measurement of $\sigma_{\text{rel}}$ are summarized in Table~\ref{syst} and explained below:
\begin{enumerate}
 \item The \Lb~and \B0~yields can vary depending on the model used to describe signal and background in data and the mass range used in the analysis. No deviations larger that 5.5\% with respect to the nominal value of $\sigma_{\text{rel}}$ are found in these tests.
 \item The cross-feed contamination between \Lb~and \B0~is quantified in simulation (see footnote~\ref{crossfeed}). Due to this effect, the result on $\sigma_{\text{rel}}$ is estimated to change at most by 2.3\%.
 \item The relative efficiency $\epsilon_{rel}$ depends on the models used in the simulation to decay the \Lb~and \B0~particles. 
  \begin{enumerate}[a)] 
   \item For \B0, the \textit{SVSCP} (scalar-vector-scalar with $CP$ violation) method~\cite{EvtGen} in  {\sc evtgen} is used, resulting in a  $2.0$\%  deviation in  $\sigma_{\text{rel}}$.
   \item The \Lb~polarization can have a large effect on the \Lb~branching fraction. Since this is the dominant systematic uncertainty, we dedicate the following subsection to describe this phenomenon.
   \end{enumerate}
\end{enumerate}

\begin{table}[h]
\begin{center}
\caption{Systematics uncertainties on $\sigma_{\text{rel}}$.}
\begin{tabular}{|l|c|}\hline
\textbf{Source} & \textbf{Error (\%)}\\\hline
Fit models & 5.5\\
Cross-feed contamination & 2.3\\
\B0~simulation & 2.0\\
\Lb~simulation (polarization) & 7.2\\\hline
Total (in quadrature) & 9.6\\\hline
\end{tabular}
\label{syst}
\end{center}
\end{table}

 \subsubsection{\Lb~polarization} \label{subsubsec:Lbpol}

Monte Carlo events with \Lb~initially polarized are generated following the methods used in~\cite{PolEd}. The {\sc evtgen} class \verb+HELAMP+~\cite{EvtGen} was extended to accept one additional parameter that sets the value of the \Lb~polarization, $P_b$. The polarization vector, given by
\begin{linenomath*}
\begin{equation}\label{eq:Pvector}
\vec P = \frac{\hat z \times \vec p }{ | \hat z \times \vec p | } P_b ,
\end{equation} 
\end{linenomath*}
is set to the \Lb~particle through the spin density matrix
\begin{linenomath*}
\begin{equation}
\mbox{\boldmath{$\rho$}} = \frac{1}{2}\left( \mathbf{I}+ \vec \sigma \cdot \vec P \right).
\end{equation} 
\end{linenomath*}
The momentum $\vec p$ of the \Lb~particle is defined in the lab system, $\vec \sigma$ are the Pauli Matrices and $\hat z$ is the direction of the proton beam. The \verb+HELAMP+ method  decays $\Lambda_b \rightarrow J/\psi (\mu^+ \mu^-) \Lambda (p \pi^-)$ according to the four complex  helicity amplitudes, $a_\pm \equiv  \mathcal{M}_{\pm\frac{1}{2}, 0}$ and $b_\pm \equiv  \mathcal{M}_{\mp\frac{1}{2}, -1}$, 
where $\mathcal{M}_{\lambda, \lambda'}$ denotes the amplitude for the \Lb~to decay into \L0~and \jpsi~with helicities $\lambda$ and $\lambda'$. The decay angular distribution
depends on the angles $\vec{\theta} = (\theta,\theta_1,\theta_2,\phi_1,\phi_2)$ depicted in Fig.~\ref{fig:polplots}a~\cite{Polth}.
By integrating in four angles, it can be shown that only the $\theta$ and $\phi_1$ distributions depend on $P_b$ ($\theta$ being the most relevant). In particular $\theta$ follows the relation (see  Fig.~\ref{fig:polplots}b),
\begin{linenomath*}
\begin{equation}
w(\theta\; ; a_\pm, b_\pm,P_b) \label{eq:theta} \propto 1 + P_b \alpha_b \cos\theta  ,
\end{equation}
\end{linenomath*}
where the weak parity violating asymmetry parameter $\alpha_b$ is defined as
\begin{linenomath*}
\begin{equation}
 \alpha_b = \frac{|a_+|^2 + |b_+|^2  - |a_-|^2 - |b_-|^2}{|a_+|^2 + |b_+|^2  + |a_-|^2 + |b_-|^2}.
\end{equation}
\end{linenomath*}

Helicity amplitudes and polarization are independent unknown parameters which are varied to study the effect on the \Lb~reconstruction efficiency. In particular, the slope of the $\cos\theta$ distribution ($\alpha_b P_b$) is allowed to vary in the full range from -1 to 1. As expected, the largest variations are found in the extreme cases $\alpha_b P_b = \pm 1$, resulting in a 7.2\% (maximum) deviation with respect to the nominal value of  $\sigma_{\text{rel}}$.

\begin{figure}
\includegraphics[scale=0.41]{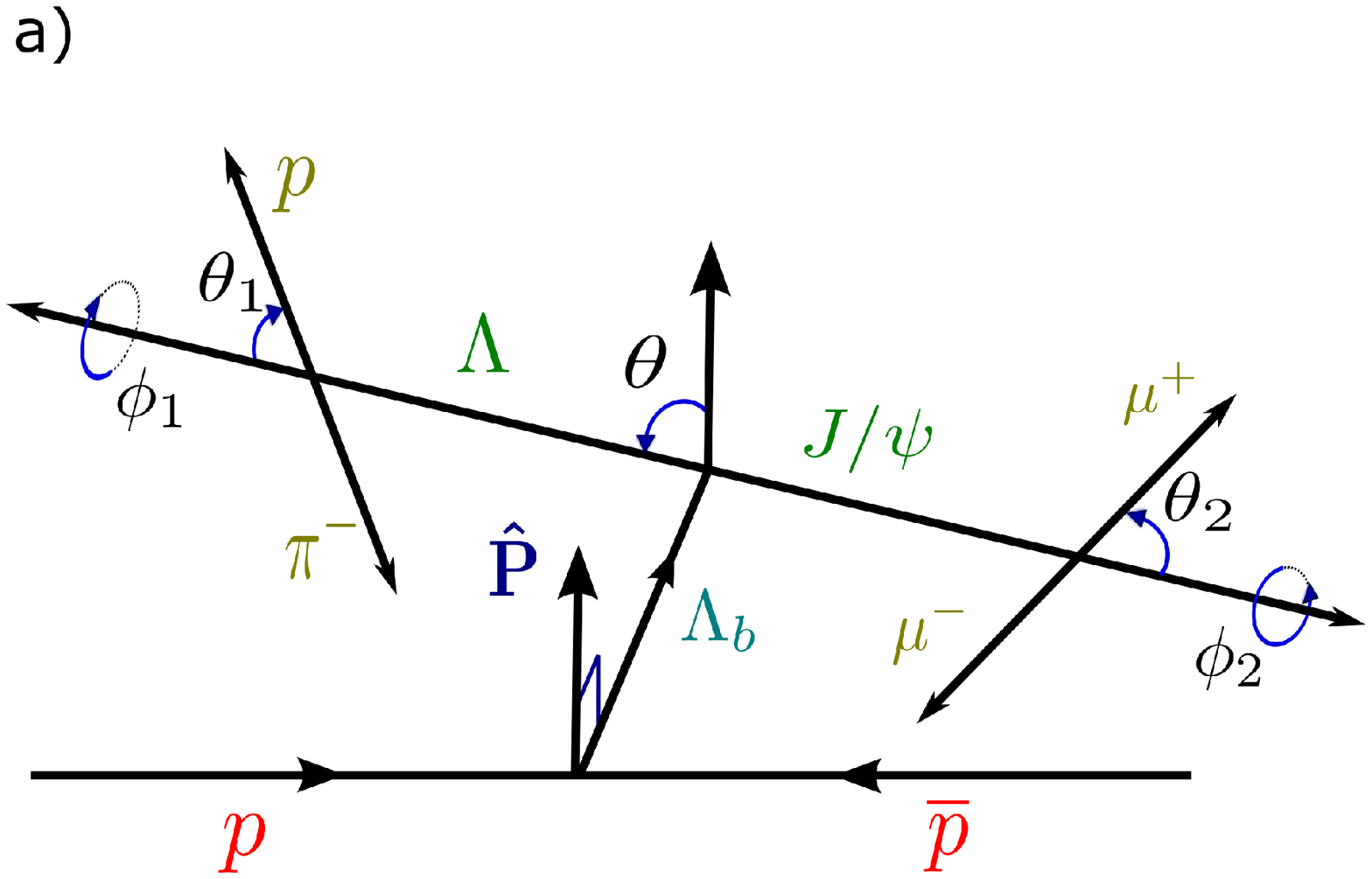} \hspace{0.2cm}
\includegraphics[scale=0.4]{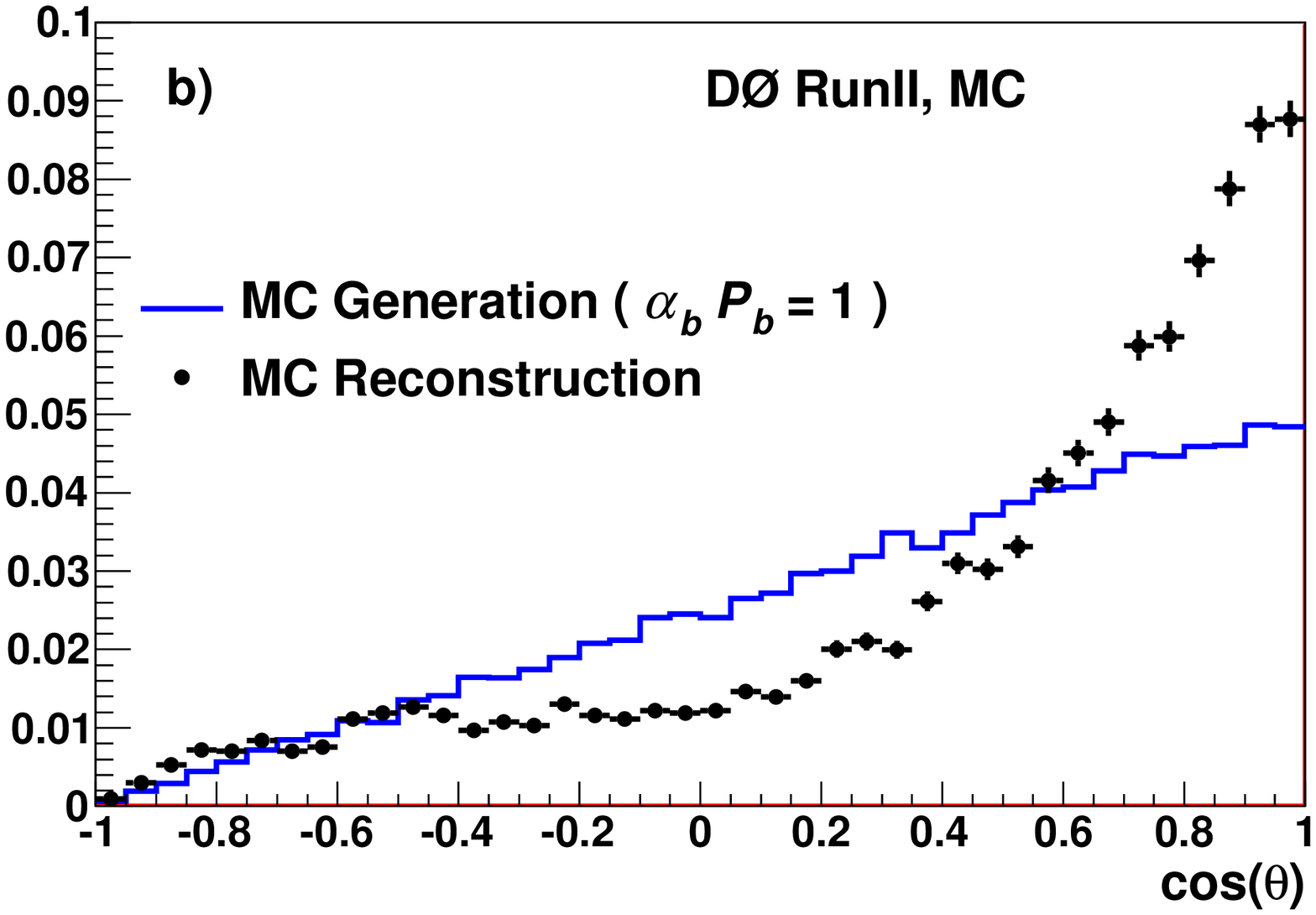}
\caption{\label{fig:polplots} a) $\Lambda_b \rightarrow J/\psi (\mu^+ \mu^-) \Lambda (p \pi^-)$ decay angles and b) $\cos\theta$ distribution of initially polarized  \Lb, with $\alpha_b P_b =1$.
}
\end{figure}

\section{Summary and discussion}
The relative production fraction times branching fraction of the \Lbch~to \Bch~decays was measured using an integrated luminosity of 6.1 \ifb~collected with the \D0 experiment. The uncertainties in Eq.~(\ref{eq:sigmarelmed}) can be combined in quadrature and the result, $\sigma_{\text{rel}} = 0.345 \pm 0.047$, can be compared with Eq.~(\ref{eq:sigmarelWA}). The error is about 3 times smaller than in the previous measurement ~\cite{CDFBrLbrecent}. Equivalently, using the best value of $f(b\to B^{0})\cdot\mathcal{B}(B^{0}\to J/\psi K^{0}_{S})=(1.74\pm 0.08)\times 10^{-4}$ from the PDG~\cite{PDG},
\begin{linenomath*}
\begin{eqnarray}
\label{Brm}
f(b\to\Lambda_{b})\cdot\mathcal{B}(\Lambda_{b}\to J/\psi \Lambda) & = & \left [6.01\pm 0.60\mbox{ (stat.)}  \pm 0.58\mbox{ (syst.)} \pm  0.28 \mbox{ (PDG)} \right ] \times 10^{-5} \nonumber \\ & = & (6.01 \pm 0.88)\times 10^{-5},
\end{eqnarray}
\end{linenomath*}
which can be compared with Eq.~(\ref{eq:fBLWA}).

The branching fraction $\mathcal{B}(\Lambda_{b}\to J/\psi \Lambda)$ is slightly more difficult to report since there is not a published measurement of $f(b\to\Lambda_{b})$. On the other hand, the \D0 and CDF experiments have observed other weakly decaying baryons such as the $\Xi_b^-$, $\Xi_b^0$ and $\Omega_b^-$, so the general assumption that $f(b  \to b_{baryon}) = f(b \to \Lambda_b)$ is not correct. A better approximation is to include the contribution of the $\Xi_b$ in the calculation, such that $f(b\to b_{baryon}) \approx f(b\to\Lambda_{b}) + f(b\to\Xi_b^-) + f(b\to\Xi_b^0)$. Furthermore, we can assume isospin invariance to  set $f(b\to \Xi_b^-) = f(b\to \Xi_b^0)$. It was also observed in Ref.~\cite{D0Cascade} that $f(b \to B_s)/ f(b \to B^0) \approx f(b \to \Xi_b^-)/ f(b \to \Lambda_{b})$. Using the PDG values of $f(b \to B^0)$, $f(b \to B^0)$ and  $f(b\to b_{baryon})$ (from the combination of LEP and Tevatron results) and their correlations~\cite{PDG}, we obtain 
\begin{linenomath*}
\begin{eqnarray}
 \mathcal{B}(\Lambda_{b}\to J/\psi \Lambda) & \approx & \frac{f(b \to B^0)}{f(b  \to b_{baryon})} \times \left[1+ 2\frac{f(b \to B_s)}{f(b \to B^0)}\right]\times \mathcal{B}(B^0 \to J/\psi K^0_S) \times \sigma_{\text{rel}} \nonumber \\
 & = & \left [11.08 \pm 1.09 \text{ (stat.)} \pm 1.06 \text{ (syst.)} \pm 3.13 \text{ (PDG)} \right] \times 10^{-4}  \nonumber \\
 & = & (11.08 \pm 3.48 ) \times 10^{-4}.
\end{eqnarray}
\end{linenomath*}

The same assumptions on $\sigma_{\text{rel}}^{W.A.}$ leads to $\mathcal{B}(\Lambda_{b}\to J/\psi \Lambda) = (8.67 \pm 4.84) \times 10^{-4}$. Both results are consistent within errors and favor theoretical models which predict a larger value for this branching ratio (see section~\ref{sec:ExpTheo}).

One final (but not less important) comment is that these measurements are useful to study $b\to s$ decays such as $\Lambda_{b}\to \mu^{+} \mu^{-} \Lambda$. Due to their similar decay topology, \Lbch~can be used to normalize  $\Lambda_{b} \rightarrow \mu^+ \mu^- \Lambda$:
\begin{linenomath*}
\begin{equation}
\mathcal{B}(\Lambda_{b} \rightarrow \mu^{+} \mu^{-}\Lambda) = \frac{N_{\Lambda_{b} \rightarrow \mu^{+} \mu^{-} \Lambda}}{ N_{\Lambda_{b} \rightarrow J/\psi  \Lambda} } \times \mathcal{B}(\Lambda_{b} \rightarrow J/\psi \Lambda) \times  \mathcal{B}(J/\psi \rightarrow \mu^{+} \mu^{-}) \times \frac{\varepsilon_{\Lambda_{b} \rightarrow J/\psi  \Lambda}}{\varepsilon_{\Lambda_{b} \rightarrow \mu^{+} \mu^{-} \Lambda}}.
\end{equation}
\end{linenomath*}
The branching fraction of this rare decay, predicted to be $\sim (2 - 5) \times 10^{-6}$ in the Standard Model~\cite{SMrare1,SMrare2,Wang}, can be enhanced by new physics effects. Recent results by CDF report the observation of this decay~\cite{CDFrare}. Using the \D0 measurement of $f(b\to\Lambda_{b})\cdot\mathcal{B}(\Lambda_{b}\to J/\psi \Lambda)$~\cite{D0BrLb}, they found $\mathcal{B}(\Lambda_{b}\to \mu^+ \mu^- \Lambda) = [ 1.73 \pm 0.42 \text{ (stat.)} \pm 0.55 \text{ (syst.)}] \times 10^{-6}$ and no significant deviation from the Standard Model.

\bigskip 

\end{document}